\documentclass[a4paper,11pt]{article}
\usepackage{pos}
\usepackage{graphicx}

\newcommand{\im}{{\rm Im}}
\newcommand{\re}{{\rm Re}}
\newcommand{\be}{\begin{eqnarray}}
\newcommand{\ee}{\end{eqnarray}}
\newcommand{\ba}{\begin{array}}
\newcommand{\ea}{\end{array}}
\newcommand{\bi}{\begin{itemize}}
\newcommand{\ei}{\end{itemize}}

\title{Non-diagonal DVCS and studies of hadronic structure with $N \to N^*$
 transition GPDs}
\ShortTitle{Non-diagonal DVCS and studies of hadronic structure}

\author*[a,b]{Kirill~Semenov-Tian-Shansky}

\affiliation[a]{Kyungpook National University, \\
Daegu 41566, South Korea}

\affiliation[b]{NRC ``Kurchatov Institute''—PNPI,\\
Gatchina 188300, Russia}

\emailAdd{ksemenov@knu.ac.kr}

\abstract{Transition generalized parton distributions (GPDs) describe matrix elements of nonlocal partonic QCD operators between the ground and excited baryon states and provide new tools for quantifying and interpreting the structure of baryon resonances in QCD. We discuss a description of non-diagonal Deeply Virtual Compton Scattering process involving a transition between a nucleon and a nucleon resonance in the
$\pi N$ system within the framework of transition GPDs. We address the physical content of $N \to N^*$
 and $N \to \Delta$
 transition GPDs, review the existing theoretical models and present theoretical estimates of related observables for the kinematic conditions corresponding to the experimental studies with JLab@12GeV. We also discuss the perspective of exploring resonance production with help of transition GPDs and consider the application of the Froissart-Gribov projections to study excitation of nucleon resonances by means of QCD probes with spin-$J$.}

\FullConference{The 21st International Conference on Hadron Spectroscopy and Structure (HADRON2025)\\
 27 - 31 March, 2025\\
Osaka University, Japan\\}

\begin{document}
\maketitle

\section{Introduction}

Generalized Parton Distributions (GPDs), see {\it e.g.} Refs.~\cite{Goeke:2001tz,Lorce:2025aqp} for an overview,  can be accessed through studies of hard exclusive processes, such as the deeply virtual Compton scattering (DVCS) and hard exclusive electroproduction of mesons (DVMP). This offers a powerful framework for exploring the internal structure of hadrons in terms of the fundamental degrees of freedom of QCD. The possibility to mimic a spin $J=2$ graviton probe due to the non-local nature of the hard subprocesses of DVCS/DVMP brings access to hadronic matrix elements of the
QCD energy-momentum tensor (EMT) known as gravitational form factors (FFs) of hadrons. This, in turn, provides a mean to investigate the decomposition of the nucleon spin and to probe
the mechanical properties of hadronic medium. Furthermore, the GPD formalism allows for construction of three-dimensional tomographic images
of hadrons, combining information in both longitudinal momentum and transverse spatial coordinates to map the distribution of quarks and gluons inside hadrons. GPDs have been in focus of intensive studies during the last three decades, see {\it e.g.} Ref.~\cite{Diehl:2023nmm}.

Recent progress has enabled the extension of the GPDs framework to include transitions involving excited nucleon states.
This offers a novel approach for probing the internal structure of nucleon resonances. On the theoretical side, the notion of transition GPDs for processes, such as
$N \to \Delta$
and
$N \to N^*$
\cite{Semenov-Tian-Shansky:2023bsy},
has been introduced, allowing the application of quark (and gluon) tomography to excited nucleons. On the experimental side, initial measurements of exclusive reactions involving
$N \to \Delta,\,N^*$
transitions have been successfully carried out at JLab
\cite{CLAS:2023akb}, confirming feasibility of such studies. An overview of the research program for investigating the structure of excited baryons
within the transition GPDs framework is presented in the white paper \cite{Diehl:2024bmd}.

\section{Non-diagonal DVCS: motivation and physical contents}

\begin{figure*}[h]
\centering
\includegraphics[width=0.9\textwidth]{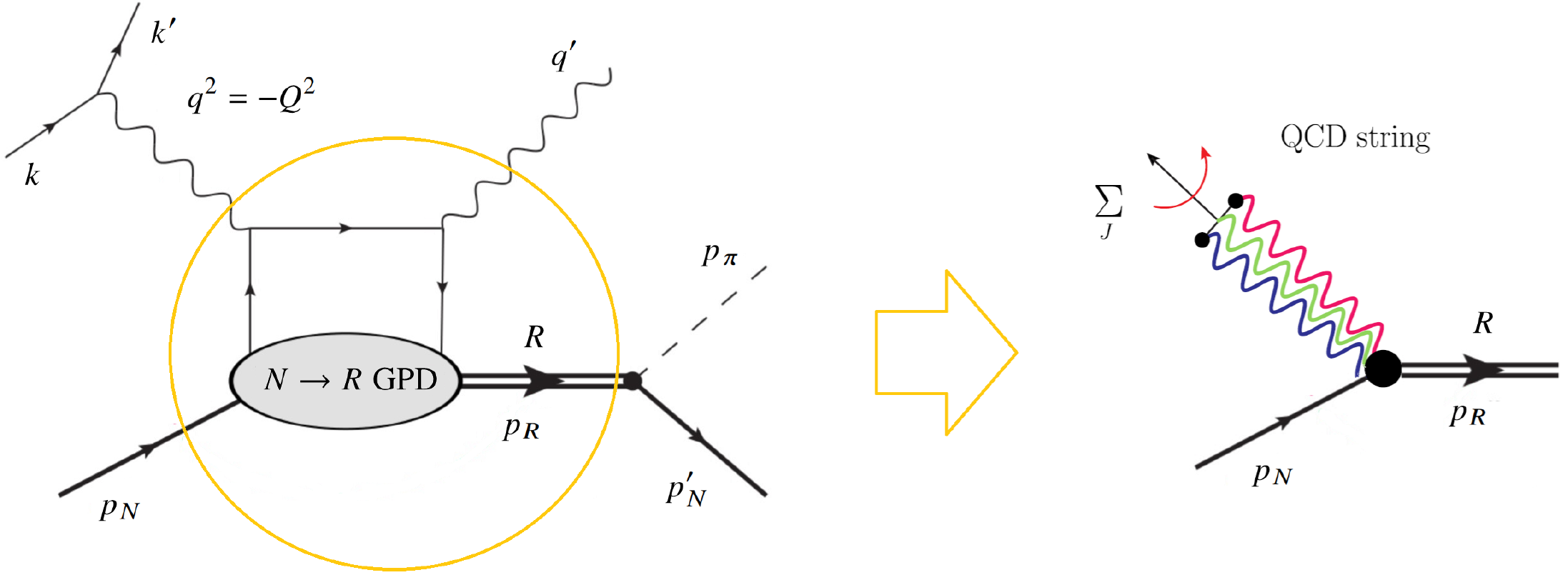}
\caption[]{Non-diagonal DVCS involving a transition from nucleon to an excited nucleon state $R=\Delta, \,N^*$. Corresponding hard subprocess may be seen as
excitation of a nucleon resonance by a non-local QCD string operator (\ref{QCD_string}), which can be expanded into a tower of local probes of arbitrary spin-$J$.}
\label{fig:NDDVCS}
\end{figure*}

Non-diagonal DVCS involving a $N \to R$ transition with $R= \Delta,\,N^*$ occurs as the hard
subprocess 
$\gamma^*(q)+ N(p_N) \to R(p_R)+ \gamma(q')$
of the hard exclusive reaction
\be
\begin{aligned}
e^{-}(k)+N(p_N) \rightarrow e^{-}\left(k'\right)+\gamma\left(q'\right)+R\left(p_{R}\right)
 \rightarrow e^{-}\left(k'\right)+\gamma\left(q^{\prime}\right)+\pi\left(p_\pi\right)+N\left(p^{\prime}_N\right),
 \label{Def_NDDVCS}
\end{aligned}
\ee
see Fig.~\ref{fig:NDDVCS}.
 The collinear factorized description in terms of nucleon-to-resonance transition GPDs
applies for this reaction in the generalized Bjorken limit characterized by large $Q^2 \equiv -q^2$ and $s=(p_N+q)^2$, with fixed $x_B \equiv  {Q^2}/{(2 p_N \cdot q)}$.
The invariants $t=(q-q')^2$, $t'=(p_N-p'_N)^2$ and the invariant mass of the $\pi N$ system,  $M_{\pi N}^2=(p'_N+p_\pi)^2$, are supposed to be of hadronic mass scale, while the invariant mass of the
real-photon-pion system,
$M_{\pi \gamma}^2 = (q'+p_\pi)^2$, is of order of the hard scale $Q^2$.

Nucleon-to-resonance transition GPDs are defined through non-diagonal matrix elements of
non-local QCD quark and gluon operators on the light-cone ($z^2=0$):
\be
\langle R| \bar{\psi}(0)[0 ; z] \psi(z)|N\rangle, \ \ \ \ \langle R| G_{\alpha \beta}^a(0)[0, z]^{a b} G_{\mu \nu}^b(z)|N\rangle,
\label{QCD_string}
\ee
where and $[0; z]$ stand for the corresponding Wilson lines.
It is worth mentioning that, although non-diagonal DVCS is a hard process, it probes a soft excitation of $R$
by a low energy QCD string, since the invariant momentum transfer from the QCD string to the target nucleon, $t=(q_1-q_2)^2$, is of
hadronic mass scale. Therefore, in some aspects, the non-diagonal DVCS reaction is more analogous to  $R$ photoproduction, rather than $R$ hard electroproduction.
The non-local QCD string operators (\ref{QCD_string}) can be expanded into towers of local probes:
$J=1$ (photons); $J=2$ (gravitons) {\it etc}. This provides access to transition
electromagnetic, gravitational and higher-$J$ transition FFs.

One of the interesting opportunities associated with non-diagonal DVCS (and DVMP)  reactions turns  to be
baryon spectroscopy studies through excitation of baryon resonances by means of non-local QCD probes.
The major advantage consists in exciting of resonances by arbitrary spin-$J$ probes with the complete
control of the QCD structure of such probes. Moreover, through non-diagonal DVMP one can get a direct access to gluonic
degrees of freedom. This can bring unique tool to study possible exotic states which couple ``weakly'' to
conventional elementary probes ($\gamma$, $W^\pm$, $Z^0$). 
It may also provide new insights into the long-standing question 
\cite{Owa:2025mep}
of the nature of the Roper resonance, ${P}_{11}(1440)$.

Generally, the observables linear in the amplitude, such as the beam spin asymmetry (BSA) $\sim \im {\cal A}$, and the 
beam charge asymmetry (BCA) $\sim \re {\cal A}$,
may considerably simplify the application of the partial wave (PW) analysis methods \cite{Arndt:2006bf}.
In 2008, a dedicated spectroscopy program 
\cite{Bspec}
was proposed for JLab. More than 15 years later,
with the arrival of data from the JLab@12~GeV, this program has become increasingly
feasible, and even greater potential is anticipated with a possible upgrade to 20~GeV.

Let us now come to the discussion of hadronic imaging with transition GPDs. The most desired is some sort of
probabilistic interpretation for images within the impact parameter representation of the Compton FFs.
However, even with the usual electromagnetic FFs that kind of perspective is somewhat tricky.
As explained in Ref.~\cite{Carlson:2007xd},
the interpretation of FFs as quark transverse charge densities is achieved in the infinite momentum frame. In case of
non-diagonal processes, the things are less obvious, because the initial and final states are not the same. So what is the  way out and
how to understand the ``images'' we get?

In order to give some idea of the physical contents of transition form factors, let us discuss the case of
$N \to \Delta(1232)$ transition. There turn to be 3 FFs: Magnetic, Electric and Coulomb quadruple.
The possibility to establish a correspondence between transitional FFs and quantities, admitting a
probabilistic interpretation is based on the SU$(2N_f)$ spin-flavour symmetry, that turns to be the
symmetry of QCD in the limit of infinite number of colors $N_c$. Both the nucleon and $\Delta$ belong to
$56$-dimensional multiplet of the SU$(6)$. This implies simple group-theoretical
relations between the form factors:
$$
G_M(0)= \frac{2 \sqrt{2}}{3} \mu_p; \ \ \ G_E(t), \; G_C(t) \sim {\rm SU}(6) \; \text{breaking effects},
$$
where $\mu_p$ is the magnetic moment of a proton. 
Non-zero $G_E$ and $G_C$ signal deviation from the $SU(6)$ symmetry
caused by quark mass differences, relativistic effects, as well as long-range QCD effects (pion cloud).
These aspects illuminate the plots of Ref.~\cite{Carlson:2007xd} for the quark transverse $N \to \Delta$  transition charge densities and
can be straightforwardly generalized for the case of $N  \to \Delta$ transition GPDs. In brief, the corresponding images are not density plots of
some charge densities, but still admit a physical interpretation as a visual reflection of symmetry breaking effects.
The emerging picture can be enriched by systematical calculation of $1/N_c$ corrections, and through implementing the information from lattice QCD.
These ideas were recently applied in Ref.~\cite{Kim:2023xvw} for the case of $N \to \Delta$ transition matrix elements of the QCD EMT introducing the
concept of QCD angular momentum for transitions between baryon states.

\section{Non-diagonal DVCS in first and second resonance regions}

Experiment can provide us with the 7-fold cross section 
$\frac{d^7 \sigma}{d Q^2 d x_B d t d \Phi d M_{\pi N}^2 d \Omega_\pi^*}$
of the $\pi \gamma$-electroproduction reaction (\ref{Def_NDDVCS}). It obtains contributions from the non-diagonal DVCS, the Bethe-Heitler (BH) process, with the 
final state photon emitted from leptonic lines, as well as the interference of the BH and non-diagonal DVCS. 
Among the 7 variables, 2 ($Q^2$ and $x_B$) are reserved to describe the lepton side of the process; 2 ($t$ and $\Phi$, the angle between leptonic and hadronic planes) 
characterize the hard stage subprocess, and, finally, $M^2_{\pi N}$ and 2 other variables describe the decay of the excited baryon into the $\pi N$ system.
For the latter variables, the convenient choice are the so-called helicity frame decay angles ($\theta_\pi^*$, $\phi_\pi^*$) defined in the rest frame of the
$\pi N$ system. 
Another characteristic observable for the reaction (\ref{Def_NDDVCS}) is the beam spin asymmetry 
\be
{\rm B S A}=\frac{d \sigma^{+}-d \sigma^{-}}{d \sigma^{+}+d \sigma^{-}},
\ee
where
$d \sigma^{\pm}$
refer to the polarized cross sections with electron beam helicity~$\pm 1/2$.
The BSA is entirely produced by the non-diagonal DVCS channel and is proportional to values of transition GPDs at the cross-over line $x=\xi$,
where $\xi$ denotes the corresponding skewness variable.

Now we are going to review the existing theoretical models for non-diagonal GPDs and
discuss the cross section estimates for the kinematical conditions of  CLAS@12 GeV.
The simplest example involves a transition from a nucleon to $\Delta(1232)$. The formalism
to describe this reaction has been known for more than 20 years \cite{Goeke:2001tz}. It generally requires $4$ vector and $4$ axial-vector
transition GPDs, see \cite{Kim:2024hhd}. However, only $3$ of them are relevant in the large-$N_c$ limit.
Early analysis was performed in Ref.~\cite{Guichon:2003ah}. Recently, we revisited the process to work out
predictions for CLAS@12 GeV. The state-of-the-art phenomenological model for $N \to \Delta$ transition GPDs
accounts for 3 GPDs, relevant in the large-$N_c$ limit that are expressed in terms of nucleon GPDs. The
Vanderhaeghen-Guichon-Guidal (VGG) model was used for the nucleon GPDs $\tilde{H}$, $E$; and the cross channel pion exchange contribution
was implemented for the nucleon GPD $\tilde{E}$.

\begin{figure*}[h]
\centering
\includegraphics[width=0.45\textwidth]{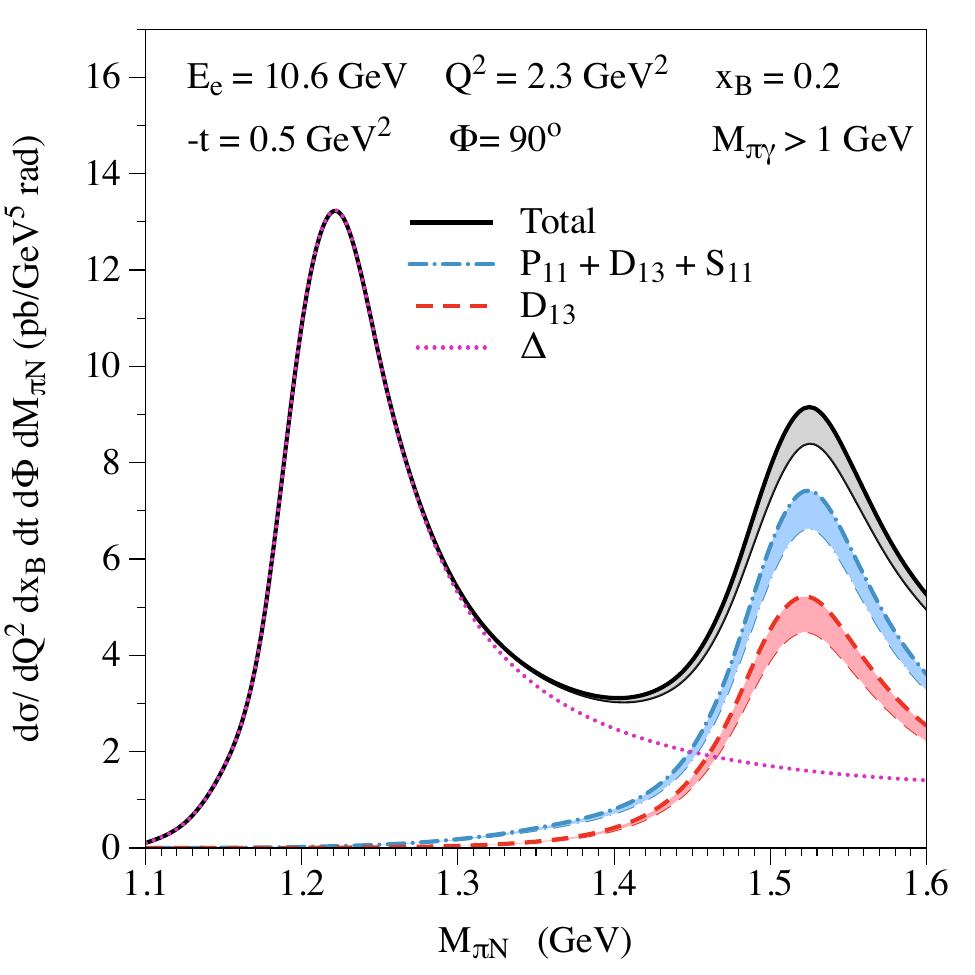}
\includegraphics[width=0.45\textwidth]{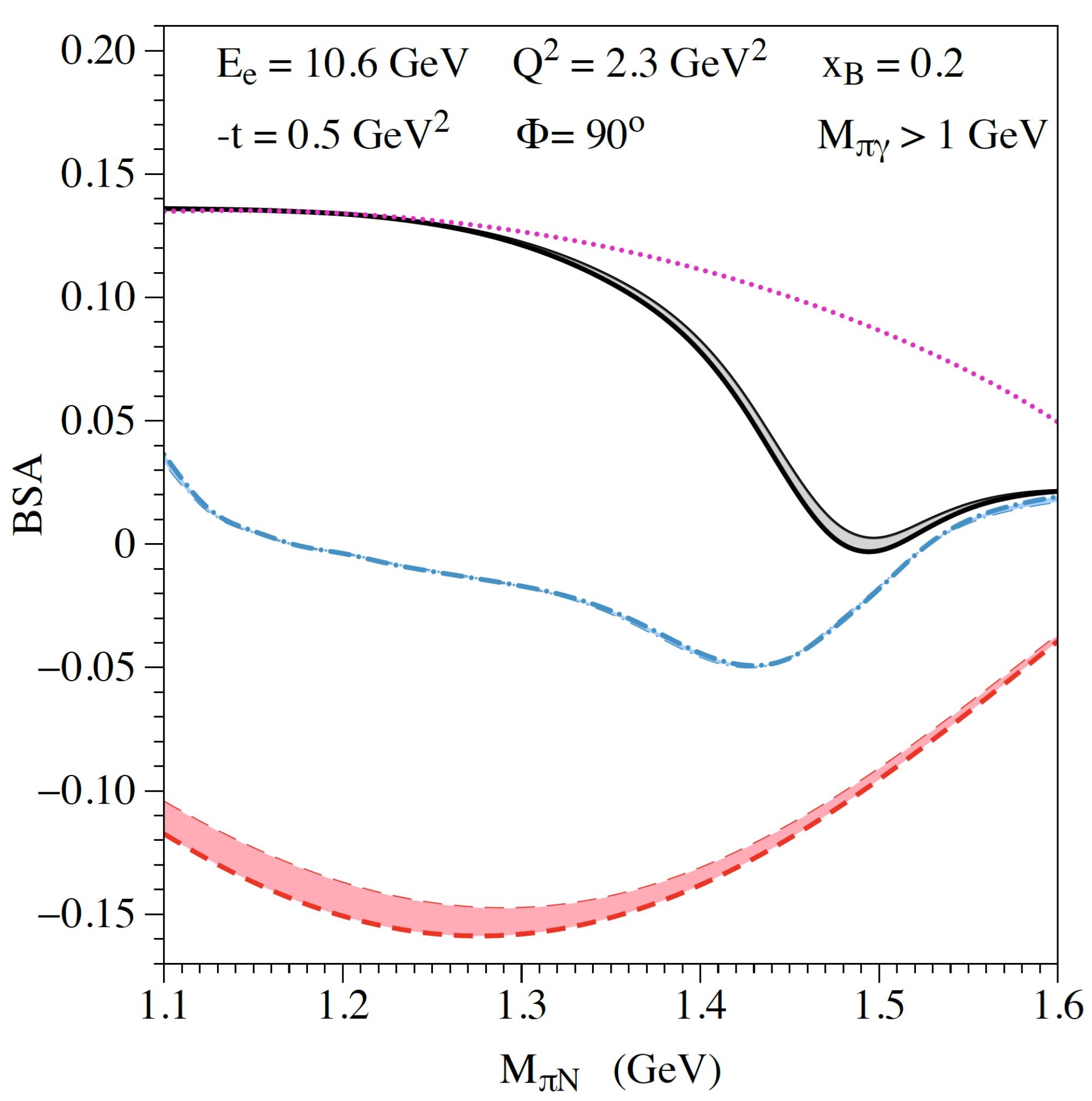}
\caption[]{Dependence on the invariant mass of the $\pi^+ n$ system ($M_{\pi N}$)
of the $e^- p \to e^- \gamma R \to e^- \gamma \pi^+ n$
cross section and corresponding BSA, integrated over the decay pion solid angle,
with the cut $M_{\pi \gamma} > 1$~GeV, for the kinematics accessible at CLAS12@JLab.
Magenta dotted curves: BH + DVCS process for $R = \Delta(1232)$;
red dashed curves and red bands: BH + DVCS process for $R = D_{13}(1520)$;
blue dashed-dotted curves and blue bands: BH + DVCS process for $R = P_{11}(1440) + D_{13}(1520) + S_{11}(1535)$;
black solid curves and grey bands: BH + DVCS process for $R = \Delta(1232) + P_{11}(1440) + D_{13}(1520) + S_{11}(1535)$. The thin (thick) curves represent the result of models I and II  implementing $x$ and $\xi$ dependence of $N \to R$ GPDs. The bands indicate the corresponding variation due to the modeling. This figure is taken from Ref.~\cite{Semenov-Tian-Shansky:2023bsy}. }
\label{fig:CS}
\end{figure*}

The step forward presented in Ref.~\cite{Semenov-Tian-Shansky:2023bsy} is the extension of the formalism to the second resonance 
region of the $\pi N$ system including the contributions of 
$N^*= {P}_{11}(1440)$, $ {D}_{13}(1520)$ and $ {S}_{11}(1535)$ resonances.
The phenomenological models for corresponding unpolarized GPDs are constrained by data on the electromagnetic transition FFs 
\cite{Ramalho:2023hqd}
measured
by CLAS@6 GeV, while polarized GPDs are governed by the Partial Conservation of the Axial Current (PCAC) principle together with  the pion pole dominance.
The $x$ and $\xi$ dependence of GPDs is introduced with help of the double distribution Ansatz with $b =\infty$ (model I) and $b=1$ (model II). 
In Fig.~\ref{fig:CS} we show the example plots for the kinematical setup typical for CLAS@12~GeV  of the 
$e^- p \to e^- \gamma R \to e^- \gamma \pi^+ n$ 
cross section (integrated over pion decay angles)
and the BSA as functions of the invariant mass of $\pi^+n$ system.

Recent analysis by the CLAS collaboration extracted the BSA for the
$\gamma^* p \to \gamma \pi^+ n$ channel as a function of $M_{n\pi^+}$ and of $-t$
in several kinematical bins, see some preliminary results in \cite{Diehl:2024bmd}. It is interesting that the phenomenological 
model of \cite{Semenov-Tian-Shansky:2023bsy} provides a reasonably good consistency with the
data, which indicates that the transition GPD approach provides an adequate description for the process.

\section{Outlook}
Let us now briefly mention further steps for development of the formalism of transition GPDs. An interesting opportunity consists in
introducing transition GPDs between a nucleon and $\pi N$ system first considered in Ref.~\cite{Polyakov:2006dd}.
$N \to \pi N$ GPDs are supposed to contain a complete information on the spectrum of the $\pi N$ system produced by means
of a non-local QCD probe in hard exclusive reactions.
The interesting feature of these objects is that they may provide a bridge between the chiral regime and the resonance regime.
In chiral regime, $N \to \pi N$ GPDs can be systematically computed with help of the chiral perturbation theory, see {\it e.g.} \cite{Alharazin:2023zzc}.
By means of the dispersive techniques, relying on the
Watson final state interaction theorem, one may extend the results towards the resonance regime.
A toy example of such connection for  spinless hadrons was recently addressed  for the $\pi \to \pi \pi$ case in Ref.~\cite{Son:2024uxa}.

Another potential development deals with a possibility to decompose the excitations induced by a QCD string into a tower of local
excitation of specific spin-$J$.
If we manage to do so, this will open new opportunities, both in spectroscopy and structural studies, as it will become possible to study hadron's
response on excitations with a given spin.

For this purpose we address the expansion of the Compton FFs in the partial waves (PWs) in the scattering angle of the cross channel process
$\gamma^* \gamma \to h \bar{h}$. The cosine of this scattering angle is in correspondence with the skewness variable $\xi$.
This type of expansion arises naturally in some approaches for GPD parametrization, including the double PW expansion in the Mellin-Barnes integral approach
and the dual parametrization of GPDs, see Ref.~\cite{Muller:2014wxa}. The Froissart-Gribov (FG) projection of the Compton FFs allows the calculation of
the FFs $F_J(t)$, encoding the response of a nucleon on the cross channel spin-$J$ excitation, using dispersive approach from the absorptive part of the DVCS amplitude  (given by GPDs on the cross-over line $x=\xi$).
Recently, in Ref.~\cite{Semenov-Tian-Shansky:2023ysr} we performed a dedicated study
of the FG projections for the case of diagonal DVCS revealing the possibility to discriminate between the existing phenomenological GPD models
through comparing model predictions for $F_J$s to the results extracted from the global analysis of DVCS data \cite{Moutarde:2019tqa}.

\acknowledgments
This work was supported by Basic Science Research Program through the National Research
Foundation of Korea (NRF) funded by the Ministry of Education RS-2023-00238703 and RS-2018-NR031074.

\end{document}